\documentclass[11pt,letter]{article}
\usepackage{subfigure}
\usepackage{amsmath}
\usepackage{amssymb}
\usepackage{graphics}
\usepackage[dvips]{epsfig}
\usepackage{times}
\usepackage{algorithm}
\usepackage{algorithmic}
\usepackage{tabularx}
\usepackage[margin=1in]{geometry}

\newcommand{\junk}[1]{}

\def\R{{\mathbb R}}

\def\SP{{\rm SP}}

\def\E{{\rm E}}

\def\cM{{\mathcal M}}

\def\deg{{\rm deg}}
\def\SP{{\rm SP}}

\def\Rec{{\rm Recall}}
\def\Prec{{\rm Precision}}

\begin{document}

\title{Mirroring co-evolving trees in the light of their topologies}
\author{Iman Hajirasouliha\,$^{1,\dagger}$, Alexander
  Sch\"onhuth\,$^{2,\dagger}$, David Juan\,$^{3}$\\ Alfonso Valencia\,$^{3}$ and
  S.~Cenk Sahinalp\,$^{1}$\\[2ex]
$^{1}$School of Computing Science\\ Simon Fraser University, Burnaby BC, Canada\\[0.5ex]
$^{2}$Centrum Wiskunde \& Informatica, Amsterdam, The Netherlands\\[0.5ex]
$^{3}$Structural Biology and BioComputing Programme\\ Spanish National Cancer Research Centre, Madrid, Spain\\
$^{\dagger}$Joint first authorship\\[2ex]
{\tt \{imanh,cenk\}@cs.sfu.ca}\\ {\tt as@cwi.nl}}

\maketitle

\begin{abstract}
Determining the interaction partners among protein/domain families
poses hard computational problems, in particular in the presence of
paralogous proteins.  Available approaches aim to identify interaction
partners among protein/domain families through maximizing the
similarity between trimmed versions of their phylogenetic trees.
Since maximization of any natural similarity score is computationally
difficult, many approaches employ heuristics to maximize the distance
matrices corresponding to the tree topologies in question.  In this
paper we devise an efficient deterministic algorithm which directly
maximizes the similarity between two leaf labeled trees with edge
lengths, obtaining a score-optimal alignment of the two trees in
question.

Our algorithm is significantly faster than those methods based on
distance matrix comparison: $1$ minute on a single processor vs.~$730$
hours on a supercomputer.  Furthermore we outperform the current
state-of-the-art exhaustive search approach in terms of precision as
well as a recently suggested overall performance measure for
mirrortree approaches, while incurring acceptable losses in recall.

\begin{sloppypar}
A C implementation of the method demonstrated in this paper is
available at http://compbio.cs.sfu.ca/mirrort.htm
\end{sloppypar}
\end{abstract}

\section{Introduction}
\label{sec.intro}

The vast majority of cellular functions are exerted by (combinations
of) interacting gene products.  As a result, "preservation of
functionality" among proteins and other gene products typically
implies "preservation of interactions" across species.  It is well
established that protein-protein interactions (both physical
interactions as well as co-occurence of domains) are preserved through
speciation events (see \cite{Lovell10,Pazos08} and the references
therein).  A major implication of this is that the evolutionary trees
behind two interacting protein families can look near-identical.\par

As interacting proteins have a tendency to co-evolve, it is desirable
to "measure" how similarly two or more proteins (or other gene
products) evolve to assess their possibility of being interaction
partners.  For that purpose a number of computational strategies have
been developed to compare the phylogenetic trees that represent two or
more protein or protein-domain families.  Among these strategies we
will focus on the {\em mirrortree} approach, where the phylogenetic
trees of protein or protein-domain families are called {\em gene
  trees}: here leaves represent "homologs" and internal vertexes
represent either speciation or duplication events.  There are a number
of mirrortree methods described in the literature each of which based
on a specific measure of pairwise tree similarity and an algorithm to
compute it; see the introductory paper by \cite{Pazos01} and
\cite{Pazos08} for more references.\par

In the context of mirrortree approaches, direct comparison of gene
trees is considered to be ``... a problem yet to be fully resolved.''
\cite[p.~2]{Izarzugaza08}, and thus available techniques typically
"measure" tree similarity in terms of the similarity between their
"distance matrices"; the distance matrix of a gene tree is defined so
that the entry $(i,j)$ represents the distance between vertices $i$
and $j$ on the tree.  Similarity between distance matrices of two
trees can easily be computed and may be used to accurately estimate
the similarity between the corresponding gene trees \cite{Pazos08} in
the absence of paralogous proteins.  This is due to the fact that the
absence of paralogs imply a bijection between the leaves of the two
trees compared (i.e. there is exactly one vertex for each species in
each gene tree).  In the presence of paralogous proteins, however, one
needs to determine the correct "pairing" of leaves so as to assess the
"degree" of co-evolution among the two families.  Note that it is not
trivial to establish such a mapping: as pointed out in
\cite{Tillier06}, protein interaction can be preserved during
duplication, while interaction can be lost during speciation.

There are a number of mirrortree approaches for determining the exact
correspondence between the leaves of two gene trees; typically these
approaches aim to "align" the distance matrices by shuffling and
eliminating the rows (and corresponding columns) so as to maximize the
similarity between the matrices.\footnote{ Note that mirrortree
  approaches differ from approaches that aim to reconcile evolutionary
  trees into a single summary \cite{Page94}.}  The similarity between
two aligned matrices is defined in the form of root mean square
difference \cite{Ramani03}, correlation coefficient \cite{Gertz03},
information-theoretic 'total interdependency' of multiple alignments
\cite{Tillier06}, Student's $t$ \cite{Izarzugaza08} or the size of
the largest common submatrix \cite{Tillier09}.  Because an exact
solution to the matrix alignment problem (where the goal is to
maximize any of these notions of similarity) is hard to compute, many
available approaches employ heuristics based on swapping pairs of
rows/columns in a greedy fashion.  These methods also commonly perform
column/row elimination from the "larger" matrix only, and not the
other \cite{Gertz03,Izarzugaza08,Jothi05,Ramani03,Tillier06}.  We are
aware of one exception by \cite{Tillier09}, which aims to determine
the largest common (i.e. within a threshold) submatrix and removes the
remainder of the columns and rows from both matrices.  Similarly the
only approach which directly compares the tree topologies themselves
is by \cite{Jothi05}, which uses a Metropolis algorithm to
heuristically travel 'tree automorphism' space.  However, this
approach can not handle trees of different sizes. See
\cite{Lovell10,Pazos08,Tillier09} for references on mirrortree
approaches which do not necessarily relate to the mapping problem.

\paragraph{Our Approach: Modeling and Formalization.}\par
In this paper we present polynomial-time algorithms that determine
mappings of leaves which respect the topology of their two gene
trees. As input, we are given two "gene trees" $T$ and $T'$ of two
protein/domain families known to interact with one another. $T$ and
$T'$ have labeled leaves where labels reflect species such that the
presence of the same label at two different leaves reflects the
presence of paralogs.  We then delete both leaves and inner vertices
from both trees until the remaining trees are isomorphic, i.e.~that is
one can map the vertices of the two remaining trees in a one-to-one
fashion onto another such that ancestor relationships are
preserved. This in particular implies a one-to-one mapping of the
remaining leaves, which we present as output. Clearly, there are many
different possible choices of such one-to-one mappings of leaves---our
algorithms determine the {\em score-optimal} such mapping where
different deletion operations are penalized in different ways,
depending on how they transform the topologies of the trees. We describe
the nature of our scoring scheme in a little more detail in the following;
please see the Methods section for full details and precise notations.

We denote a bijection (i.e. a one-to-one and onto mapping) of subsets
of vertices of $T,T'$ by 
$\cM[T, T']$ and write 
\begin{equation}
M:=\{(v,w)\in T\times T'\mid \cM(v) = w\}
\end{equation}
for the pairs of mapped vertices.  Note that in such a bijection, not
all vertices of $T$ are necessarily mapped to a vertex in $T'$ and
vice versa. We refer to vertices which are not mapped as {\em deleted}
by $\cM[T,T']$.  We only consider mappings which satisfy the
following: (1) the mapping preserves the ancestor relationship of $T$
and $T'$; (2) only leaves with identical labels are mapped onto one
another; (3) upon deletion of vertices, where deletion of an internal
vertex $v$ leads to new edges joining the parent of $v$ with the
children of $v$, the two tree topologies are isomorphic.  Among the
mappings satisfying the above conditions, we compute 
the mapping that has maximum score.

For a formal definition of our scoring scheme, consider the internal
vertices of $T$ and $T'$ that are deleted.  Among them, we distinguish
between vertices $v$ that have descendants $x$ which are not deleted.
We write $N_I$ for such vertices. We write $N_T$ for the remaining
deleted vertices. Note that each vertex $v\in N_T$ makes part of a
subtree of $T$ which has been deleted as a whole.  The score of the
mapping is then defined as
\begin{equation}
\label{eq.score1}
S(\cM[T,T'])\\ = \sum_{(v,v')\in M}S_M(v,v') + \sum_{v\in N_I}S_{N_I}(v) + \sum_{v\in N_T}S_{N_T}(v).
\end{equation}
The individual score functions $S_M, S_{N_I}$ and $S_{N_T}$ will be
formally defined in the Methods section.  Our algorithm, which
maximizes the overall score of the mapping, can be viewed as an
extension of the standard {\em tree edit distance} algorithm for
unweighted trees (e.g.~\cite{Tai79}), to those with edge weights.
Determining the tree edit distance is NP-complete~\cite{Zhang92} (in
fact MAX-SNP-hard~\cite{Zhang94}).  Since the instances treated here
are too large (trees have up to more than $200$ leaves) we have to
impose reasonable constraints when aiming at fast, polynomial-runtime
solutions. Motivated by test runs (see numbers referring to
$C_{1,2,3}$ in the Results and Discussion section), we chose to impose
the additional constraint that a vertex $u$ and its parent $v$ cannot
be deleted at the same time without that the entire subtree rooted at
$v$ is deleted. That is we disallow to have both a parent $v$ and a
child $u$ in $N_I$.  Note, however, that deletion of two internal
siblings is permissable---we found that such deletions can lead to
favorable mappings.  As the operation of deleting entire subtrees does
not lead to runtime issues, does not perturb the topology of the
remaining trees and also reflects the biologically reasonable
assumption that interaction can be lost for entire subtrees, we allow
it without additional restrictions.

Note that the algorithm only outputs one uniquely determined,
score-optimal mapping of subsets of leaves of $T,T'$. Note further
that we do not perform an exhaustive search since we do never consider
mappings of leaves which imply mappings of internal vertices that do
not preserve ancestor relationships of the gene trees $T,T'$ and
thereby contradicts their topologies.

Alternative constraints leading to polynomial time solvable variants
on the tree edit distance is surveyed in \cite{Zhang96}.  For further,
more recent work see also \cite{Pinter08} that address the subtree
homeomorphism problem, which, given a "text" tree $T$ and a "pattern
tree" $P$ as the input, asks to find a subtree $t$ in $T$ such that
$P$ is homeomorphic to $t$.  Now, two trees $T_1,T_2$ are said to be
homemorphic if one can remove degree $2$ vertices from $T_1,T_2$ such
that $T_1$ and $T_2$ are isomorphic.  Another recent work
\cite{Raynal10} considers homeomorphic alignment of "weighted" but
unlabeled trees.  Here the goal is to obtain a homeomorphic mapping
between vertices of two trees such that the differences between the
weights of "aligned" edges is minimized.  While being related to our
approach, the method described in \cite{Raynal10} is not applicable to
our problem as the trees they consider are not leaf labeled.  We refer
the reader to \cite{Bille03} for a general and gentle overview of
further related work on tree edit distance, tree alignment and tree
inclusion.

\paragraph{Summary of Contributions.}
The main technical contribution of this paper is a novel deterministic
mirrortree algorithm that directly compares tree topologies.  The
algorithm is optimal within the single constraint we impose and is
provably efficient.  We compare our algorithm with the most recent,
state-of-the-art heuristic search approach \cite{Izarzugaza08} that
aims to maximize the similarity between distance matrices, where
distances reflect lengths of shorted paths in neighbor-joining trees.
In our comparisons we use precisely the same trees to be able to
juxtapose a distance matrix-based heuristic search method to our
topology-based, deterministic method without introducing further
biases.  Our main conclusions are as follows.
\begin{itemize}
\item We can compute mappings for the $488$ interacting domain
  families in roughly $1$ minute on a single CPU - in comparison to
  $730$ hours on MareNostrum \footnote{MareNostrum is a supercomputer
    of the Barcelona Supercomputing Center, one of the largest
    machines in the world dedicated to science
    \cite[p.~10]{Izarzugaza08}.}  needed for the Metropolis search
  performed by \cite{Izarzugaza08}.
\item We outperform the Metropolis search in terms of precision, i.e.,
  the percentage of correctly inferred pairings among all inferred
  pairings is higher ($48\%$) in our approach vs.~that ($43\%$) in
  \cite{Izarzugaza08});
\item In terms of F-measure (see the Discussion section for a
  definition), which has been most recently suggested for assessing
  mirrortree approaches in terms of both recall and precision
  \cite{Tillier09}\footnote{Note that \cite{Tillier09} suggest
    $F_{0.1}$, which favors the topology-based approach even more than
    $F_{0.25}$ which we use.}, our topology-based approach again
  prevails ($0.47$ over $0.45$).
\end{itemize}

\section{Preliminaries and notations}

Let $T=(V, E, w)$ be a tree with weighted edges as given by a
non-negative weight function $w:E\to\R_+$.  We denote the leaves of
$T$ by $L=\{\ell_1,...,\ell_n$\}, the internal nodes of $T$ (excluding
the root) by $U=\{u_1,...,u_m\}$, and the root of $T$ by $r$. In
particular let $n$ be the number of leaves and $m$ be the number of
internal vertices without the root. Note that a tree $T$ is binary and
rooted if and only if $\deg(r)=2$ and $\deg(u) = 3$ for all internal
vertices $u\in U$; this will imply that $m=n-2$ and $|E|=2n-2$.  In
our setting, edge weights $w(v_i,v_j)$ reflect the evolutionary
distance between adjacent vertices $v_i,v_j$.  Note that leaves refer
to gene products whereas internal vertices can be interpreted as
speciation and/or duplication events.  For a given vertex $v \in V$,
we define $\theta(v)$ as the evolutionary distance between the root
and $v$.  In other words, $\theta(v)$ is the sum of the edge weights
in the unique path from the root to $v$.  In rooted trees, there is a
natural partial order
\begin{equation}
\label{eq.partialorder}
v_i\le v_j \quad\Leftrightarrow\quad \mbox{$v_i$ is an ancestor of $v_j$}
\end{equation}
on the vertices of $T$.  Hence, the edges have a natural orientation
and each vertex $v_i$ induces a unique subtree $T(v_i)$.  This partial
order is crucial for our algorithm---which can not be applied to
unrooted trees in a straightforward manner.  For processing unrooted
(e.g. neighbor-joining) trees, consider the pair of proteins/domains
(one from each tree) which are known to interact.  We root the two
trees at these vertices in order to apply our algorithm.  Provided
such a pair exists (which is typically the case), our algorithm
optimally aligns the trees as it does not assume any order among the
many sibling vertices. In a tree $T$ which is rooted at $r$, we call
vertex $u$ the parent of a vertex $v$ if $u$ and $v$ are connected by
an edge and $u$ is closer to $r$ than $v$. The {\em height} of a
rooted tree is defined as $\max \{d(r,\ell_i)\mid i=1,...,n\}$ where
$d(v_1,v_2)$ is the length of the shortest path between vertices $v_1$
and $v_2$ without considering edge weights, that is the maximum
(unweighted) distance of the root to a leaf.

\section{Methods}

\begin{figure}[!tpb]
\centering
\includegraphics[width=0.32\columnwidth]{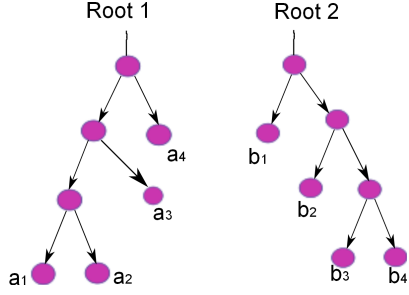}\hspace{0.05\columnwidth}
\caption{Two isomorphic trees are shown as an example in this
figure. The leaves of the left tree are labeled with $a_1, a_2, a_3,
a_4$ while the leaves of the tree on the right are labeled with
$b_1, b_2,b_3, b_4$. A possible mapping between the leaves that
respect the tree topology is $(a_1, b_3), (a_2, b_4), (a_3,
b_2), (a_4, b_1)$.}
\label{fig.isomorphic}
\end{figure}

Given two rooted weighted-edge trees $T$ and $T'$, our algorithm {\it
  aligns} the trees by mapping a subset of leaves of $T$ to a subset
of leaves of $T'$. In order to obtain this mapping, a series of (1)
individual vertex deletions or (2) subtree deletions (with specific
penalties) are performed on each tree with the goal of obtaining two
isomorphic trees $T_1=(V_1, E_1, w_1)$ (from $T$) and $T'_1=(V'_1,
E'_1, w'_1)$ (from $T'_1$); Figure \ref{fig.isomorphic} shows two such
rooted trees that are isomorphic; it also shows a mapping between the
leaves.  The specifics of vertex and subtree deletions on a tree
$T=(V,E,w)$ are as follows.
\begin{enumerate}
\item Deleting an internal vertex $v$ also deletes the edge $(u,v)$,
  where $u$ is the parent of $v$.  Furthermore, it connects each child
  $x$ of $v$ to $u$ by deleting the edge $(v,x)$ and creating a new
  edge $(u,x)$.  The weight of this new edge, $w(u,x)$ is set to
  $w(u,v)+w(v,x)$.  As mentioned earlier, it is not possible to delete
  both a node $v$ and its parent $u$ from $T$.
\item Deleting an entire subtree rooted at an internal vertex $v$
  deletes all descendants of $v$ and their associated edges.
\end{enumerate}

In the remainder of this section, we will discuss the costs of the
above deletion operations and the scores of the mapped vertices.  As
mentioned earlier, the overall score of the mapping will be the sum of
the scores of the mapped vertices and the scores (negative costs) of
the the deletion operations.

\subsection{Scoring Scheme}

Let $T_1$ and $T'_1$ be the isomorphic trees which result from
performing a series of deletion operations on $T$ and $T'$.  The
isomorphism $\Phi:T_1 \to T'_1$ implies a mapping (a.k.a. alignment)
$\cM[T,T']$ between the original trees $T, T'$.  Let $L_1,L'_1$ denote
the sets of leaves that are mapped in $T$ and $T'$ respectively;
because the mapping is a bijection, we must have $|L_1|=|L'_1|$.  We
write $\SP:=\{(l,l')\mid l\in L, l'\in L', (l,l')\in
M\}\subset\cM[T,T']$ for the set of mapped pairs (we require that
mapped leaves have identical labels hence the naming $\SP$ for
'species'.

\begin{figure}[ht]
\centering
\subfigure[No contracting edges]{
\includegraphics[scale=0.5]{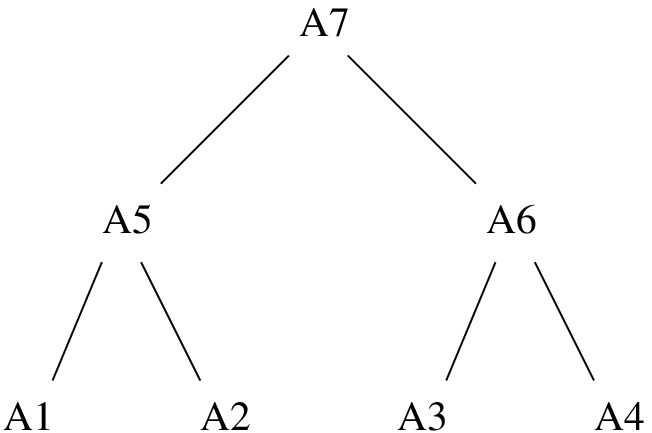} \hspace{0.35cm}
\label{fig:nocontract}
}
\subfigure[An isolated contracted edge: $(A_5, A_7)$]{
\includegraphics[scale=0.5]{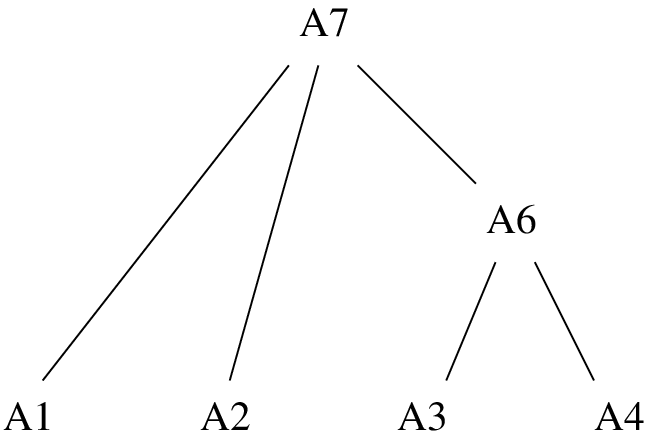} \hspace{0.35cm}
\label{fig:onecontract}
}
\subfigure[Two contracting edges: $(A_7,A_5)$ and $(A_7,A_6)$]{
\includegraphics[scale=0.5]{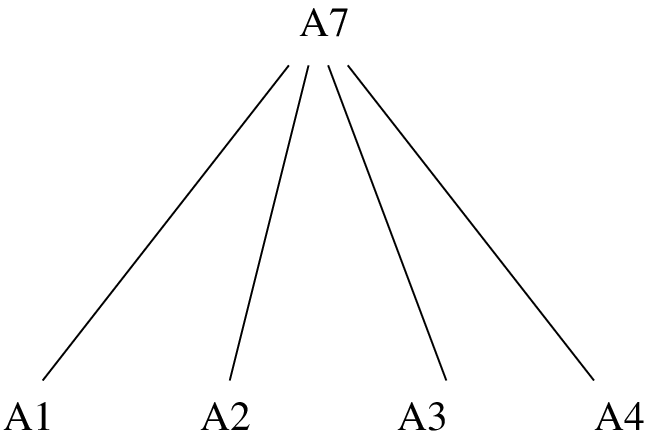}
\label{fig:twocontract}
}
\
\caption{A gene tree \subref{fig:nocontract}, with an isolated contraction of the edge $(A_5,A_7)$ subref{fig:onecontract} and a parallel contraction of the  
 edges $(A_5,A_7)$ and $(A_6,A_7)$ \subref{fig:twocontract}.
}
\label{fig:trees}
\end{figure}

Recall that a mapping of two trees may involve deleting internal
vertices or entire subtrees.  We now distinguish between two types of
internal vertex deletions, a.k.a. edge contractions.
\begin{enumerate}
\item $\,$[Isolated Deletion:] deletion of only one child $v$ of a
  vertex $u$. Let further $x_1,x_2$ be the two children of $v$.
  Isolated deletion of $v$ also implies to also delete edges $(u,v),
  (v,x_1), (v,x_2)$ and create new edges $(u,x_1),(u,x_2)$.
\item $\,$[Parallel Deletion:] deletion of both children (say $x$
  and $y$) of a vertex $v$. This implies deletion and creation of
  edges in a fashion analogous to that for isolated deletion.
\end{enumerate}
Accordingly, we further distinguish between isolated deleted vertices
$N_{I,iso}$ and vertices which became deleted in parallel $N_{I,par}$
such that $N_I=N_{I,iso}\;\dot{\cup}\;N_{I,par}$.  For a given mapping
$\cM[T,T']$ let $\E_{S}(\cM):=\{(u,v)\mid v\in
N_{I,iso}\}$ be the set of edges which join isolated deleted vertices
with their parents.  Analogously, $\E_{P}(\cM)$ is the set of edges
that join deleted siblings with their parent. See figure
\ref{fig:trees} for examples of isolated and parallel deletions.

Given a pair of mapped leaves ${\tilde{\ell}}_1, {\tilde{\ell}}_2 \in
\SP$ their alignment score, $\kappa({\tilde{\ell}}_1,
{\tilde{\ell}}_2)$ is defined as
\begin{equation*}
  \kappa({\tilde{\ell}}_1, {\tilde{\ell}}_2) = 
    C - |\theta(\tilde{\ell_1}) - \theta(\tilde{\ell_2})|\\
\end{equation*}
where $C$ is a positive constant, providing a positive contribution to
the overall score because of the alignment of two leaves with the same
label while we subtract the difference between the distances of
$\tilde{\ell_1}$ and $\tilde{\ell_2}$ from the root for penalizing the
alignment between two leaves which have topologic differences. 

The total score $\mathcal{S}$ of an alignment $\cM[T, T']$ as per the
above definition is fully specified by

\begin{equation}
\label{eq.score}
\mathcal{S}(\cM[T,T']) = \sum_{({\tilde{\ell}}_1, {\tilde{\ell}}_2) \in \SP} \kappa({\tilde{\ell}}_1, {\tilde{\ell}}_2)
 - \sum_{e_s \in E_{iso}(\cM)}E \cdot w(e_s) - \sum_{e_p \in E_{par}(\cM)}F \cdot w(e_p)
\end{equation}
\noindent where, with respect to the formulation in (\ref{eq.score1}),
the term in the first row is for $\sum_{v,v'\in M}S_M(v,v')$, the
second row is for $\sum_{v\in N_I}S_{N_I}(v)$ and $\sum_{v\in
  N_T}S_{N_T}(v)$ is zero. $E$ and $F$ are user-defined constants that
respectively penalize {\it isolated deletion} and {\it parallel
  deletion} of edges. Note that this penalty is proportional to the
length of the edges joining the deleted vertices with their
parents---deletion of longer edges leads to a more severe perturbation
of topology hence is more severely penalized. We set the cost of
deleting a subtree (i.e. $S_{N_T}$) to $0$. Note, however, deleting
subtrees is implicitly penalized by disregarding any potential good
mappings of leaves in them.

Given the above score function, the {\em gene tree alignment problem} can be formally stated as follows.
\bigskip
\hrule
\begin{center}{\bf Gene Tree Alignment Problem}\end{center} 
\vspace{-1ex} \begin{sloppypar}Given two rooted weighted-edge trees $T,T'$,
determine subsets of leaves $L_1\subset L,L_1'\subset L'$ of equal
size such that the corresponding subtrees can be transformed by
isolated and parallel edge contraction and subtree removal operations into trees
$T_1,T'_1$, for which there is an isomorphism $\Phi:T_1\to T'_1$
that maximizes $\mathcal{S}(\cM[T, T'])$.\\[-.5ex]
\end{sloppypar}
\hrule
\bigskip

\subsection{A Dynamic Programming Solution}

The gene tree alignment problem can be efficiently solved by a dynamic
 programming algorithm.  
Our algorithm runs in $O((|V|\cdot |V'|)$ time for two binary, rooted trees $T, T'$ with vertex sets
 $V, V'$. 
In general, our strategy can be applied to arbitrary rooted
 trees with bounded maximum degree, $\Delta_{\max}$. 
Note that by allowing to delete internal vertices (i.e. contract the
 edges), the number of children of
 an internal vertex will be still bounded by a constant ($\leq 4$).
\paragraph{Initialization}
\label{ssec.initial}
As a first step, we remove all leaves that refer to species that are
unique to each tree.  Let $n=|V|$ and $n'=|V'|$.  For every pair of
vertices $v_i \in V$ and $v'_j \in V'$ (i.e. for every $i=1, \cdots,
n$ and $j=1,\cdots, n'$), we compute the maximum alignment score for
the subtrees rooted at $v_i$ from $T$ (i.e. $T(v_i)$) and $v'_j$ from
$T$ (i.e. $T'(v'_j)$).  We denote the maximum alignment score for
$T(v_i$) and $T'(v'_j)$ by $S_{ij}$.

In our dynamic programming algorithm, we handle the "base" cases,
where one (or both) of $T(v_i)$ or $T(v'_j)$ have $3$ or fewer leaves,
as follows.
\begin{itemize}
\item If both $v_i \in V$ and $v'_j \in V'$ are leaves, then by
  definition, $S_{ij} = \kappa(v_i, v'_j)$.
\item Without loss of generality, if $v_i$ is a leaf and $v'_j$ is an
  internal vertex, $S_{ij} = \max (S_{ij_1}, S_{ij_2})$, where $j_1$
  and $j_2$ correspond to the children of $v'_j$.
\item The remainder of the  base cases have both $v_i$ and $v_j$ as internal vertices and are solved
through exhaustive evaluation of all possible alignments. 
\end{itemize}

\paragraph{Recursion}
\label{ssec.recursion}
internal vertices, each with at least $4$ descendants, $S_{ij}$ will
be computed through recurrence equations.  These equations are based
on the alignment scores between subtrees rooted at the children (or
grandchildren) of $v_i$ and $v'_j$.  Let $i_1(j_1)$ and $i_2(j_2)$ be
the children of the vertex $v_i(v'_j)$.  Also, let $i_{11}$, $i_{12}$
be the children of $i_1$, and $i_{21}$, $i_{22}$ be the children of
$i_2$. Similarly, let $j_{11}$, $j_{12}$ be the children of $j_1$, and
$j_{21}$, $j_{22}$ be the children of $j_2$.  We first give a high
level description of the recurrence equation.  Suppose that the
maximum alignment score between any subtree in $T(v_i)$ and any
subtree in $T'(v'_j$) has already been computed.  In order to compute
the alignment score $S_{ij}$, we consider several cases: we can either
delete one or both subtrees rooted at the children of $v_i$ and $v'_j$
(deleting an entire subtree) or align the subtrees rooted at the
children of $v_i$ and $v'_j$ to each other.  We can also delete one of
the children of $v_i$ (either $i_1$ or $i_2$) together with one of the
children of $v_j$ (either $j_1$ or $j_2$) and align the three
resulting subtrees in $T(v_i)$ to a {\it permutation} \footnote{We
  have to consider all the permutations because the trees are
  unordered (i.e. the order of siblings of an internal vertex is
  unimportant).} of the ones in $T'(v'_j)$.  Finally, we have to
consider the case where both children of the root (i.e. $i_1$ and
$i_2$ in $T(v_i)$, and $j_1$ and $j_2$ in $T'(v'_j)$) are deleted.  In
this case we align four subtrees in $T(v_i)$ (rooted at $i_{11},
i_{12}, i_{21}$, $i_{22}$) to a permutation of the four resulting
subtrees in $T'(v'_j)$. The optimal alignment score of $S_{ij}$ will
thus be the {\it maximum} alignment score provided by all of the cases
above.

Let $e(v)$ denote the penalty for isolated deletion of an internal
vertex $v$, which is the product of the constant $E$ and the weight of
the edge between $v$ and its parent (see Scoring Scheme section).
Also, let $f(v)$ denote the penalty for parallel deletion of both
children of an internal vertex $v$. $f(v)$ was defined as a constant
$F$ times the total weight of the edges that connect $v$ to its
children.  The recurrence equation for $S_{ij}$ thus becomes the
following

\begin{equation}
\label{eq.recursion}
S_{ij} = \max \begin{cases} 
	0 \text{ (deleting both subtrees from each tree)}\\             
             \begin{Bmatrix}
               S_{i_1j_1} + S_{i_2j_2}\\
               S_{i_1j_2} + S_{i_2j_1}
             \end{Bmatrix} \text{regular cases}\\               
             \begin{Bmatrix}
               S_{ij_1}\\
               S_{ij_2}\\
               S_{i_1j}\\
               S_{i_2j}
             \end{Bmatrix} 
\text{deleting one subtree from each tree}\\
    \max\begin{Bmatrix}
              S_{i_{11}j_2} + S_{i_{12}j_{11}}+S_{i_2j_{12}}\\
              S_{i_{11}j_2} + S_{i_{12}j_{12}}+S_{i_2j_{11}}\\
              S_{i_{12}j_2} + S_{i_{11}j_{11}}+S_{i_2j_{12}}\\
              S_{i_{12}j_2} + S_{i_{11}j_{12}}+S_{i_2j_{11}}
	\end{Bmatrix} - e(i_1) - e(j_1) \\ 
    \max \begin{Bmatrix}
              S_{i_{21}j_2} + S_{i_{22}j_{11}}+S_{i_1j_{12}}\\
              S_{i_{21}j_2} + S^{i_{22}j_{12}}+S_{i_1j_{11}}\\
              S{i_{22}j_2} + S^{i_{21}j_{11}}+S_{i_1j_{12}}\\
              S_{i_{22}j_2} + S_{i_{21}j_{12}}+S_{i_1j_{11}}
	\end{Bmatrix} - e(i_2)-e(j_1) \\
    \max \begin{Bmatrix}
              S_{i_{21}j_1} + S_{i_{22}j_{21}}+S_{i_1j_{22}}\\
              S_{i_{21}j_1} + S_{i_{22}j_{22}}+S_{i_1j_{21}}\\
              S_{i_{22}j_1} + S_{i_{21}j_{21}}+S_{i_1j_{22}}\\
              S_{i_{22}j_1} + S_{i_{21}j_{22}}+S_{i_1j_{21}}
\end{Bmatrix} - e(i_2)-e(j_2) \\
 \max \begin{Bmatrix}
              S_{i_{11}j_1} + S_{i_{12}j_{21}}+S_{i_2j_{22}}\\
              S_{i_{11}j_1} + S_{i_{12}j_{22}}+S_{i_2j_{21}}\\
              S_{i_{12}j_1} + S_{i_{11}j_{21}}+S_{i_2j_{22}}\\
              S_{i_{12}j_1} + S_{i_{11}j_{22}}+S_{i_2j_{21}}\\
	\end{Bmatrix} - e(i_1)-e(j_2) \\  
      S_{i_{11}\pi_1} + S_{i_{12}\pi_2} + S_{i_{21}\pi_3} + S_{i_{22}\pi_4} - f(v_i)-f(v_j)
       \end{cases}
\end{equation}

\noindent where the {\it permutation} $\pi=\pi_1 \pi_2 \pi_3 \pi_4$
ranges over all permutations of $\{j_{11},j_{12},j_{21}, j_{22}\}$.
Note that some cases are redundant but are still represented here for
the sake of clarity.

Now, given $r$ and $r'$, the roots of $T$ and $T'$, respectively, the
alignment score ${S_{r_1r_2}}$ (i.e. the maximum alignment score of
the rooted trees) can be computed using the above recurrence equation,
providing a solution to the gene tree alignment problem.  It is quite
straightforward to prove that our algorithm correctly computes the
maximum alignment score through a (strong) induction on the {\it sum}
of the heights of the rooted trees.  Note that the scores of internal
vertex alignments can be computed through the scores of the alignments
between their (grand)children and the recurrence precisely serves to
satisfy the constraints.  The base of the induction is trivial.  If
the minimum height of the trees is zero (i.e. one of the trees is just
a single leaf), the optimal value of the alignment can be found using
the definitions and simple case analysis.  Given the subtrees $T(v_i)$
and $T'(v'_j)$, with heights $h$ and $h'$, respectively, we assume the
induction hypothesis, that for all pairs of subtrees $T(v_p)$ and
$T'(v'_q)$ with heights $h_p$ and $h_q$ such that $ h_p + h_q < h +
h'$.  It is easy to verify by case analysis that all cases in the
recurrence equation will be reduced to a case in which the sum of the
heights of the aligned (grand) children will be less.

\section{Results}
\newcolumntype{C}{>{\centering\arraybackslash}X} %

\paragraph{Data Source and Alternative Methods.} We benchmarked our
algorithm against the most recent heuristic search method
\cite{Izarzugaza08} for determining a mapping in the presence of
paralogs on the large-scale data corpus described in the same study.
This data set contains multiple alignments for 604 yeast protein
domains among which 488 domain pairs are known to co-occur in the same
protein.
those 488 domain family pairs is considered to be a particularly tough
test \cite{Izarzugaza08} due to the presence of approximately $6$
paralogs per species on average.
For all interacting domain family pairs, neighbor-joining trees were
computed, using ClustalW \cite{ClustAlW} and the trees were rooted at
the domains which are known to interact.

\paragraph{Evaluation Criteria.} 
Following \cite{Izarzugaza08}, we determine the maximum number of
protein domains that can be paired without topology constraints;
i.e. if we have $k_1$ paralogs of a particular protein domain $d_1$
and $k_2$ paralogs of domain $d_2$ within the same species, then this
species contributes $\min(k_1, k_2)$ to the overall count.  By the
usual conventions, we denote this value as $P$.  Among $P$ potentially
correctly paired protein domains, the number of those which have been
inferred by the algorithm in use, such that both domains reside in the
same protein, is referred to as "true positives", $TP$.  Similarly the
number of protein domain pairings computed, which do not reside in the
same protein are determined as "false positives, $FP$.  Recall
(Sensitivity) is defined as $Rec=TP/P$ and Precision (Positive
Prediction Rate) is defined as $Prec=TP/(TP+FP)$ while the F-Measure
$F_{0.25}$\footnote{Note that $F_{0.1}$, which has been recently
  suggested as an appropriate prediction quality measure for
  mirrortree approaches \cite{Tillier09}, yields results which are
  even more in favor of our approach.}  is determined as
$(1+0.25^2)\cdot Rec \cdot Prec/(0.25^2Prec + Rec)$.  Note that Recall
is referred to as Accuracy in \cite{Izarzugaza08}. We determine
Precision, Recall and $F_{0.25}$ for each pairs of trees
individually. Values displayed in tables~\ref{tab.overall},
\ref{tab.maxsize} and \ref{tab.space} are average values for all $488$
co-evolving tree pairs.

\paragraph{Tree Constraints.} In order to appropriately assess the
contribution of the different tree constraints as outlined in the
Methods section, we evaluated our algorithm by not allowing to
contract edges ($C_0: C=1,E=\infty,F=\infty$ in
(\ref{eq.score})), allowing edge contraction (without penalty, that is
$E=0$ in (\ref{eq.score})) up to creating ternary, internal vertices
($C_1: C=1,E=0,F=\infty$ in (\ref{eq.score})) as
well as further allowing creation of quarternary vertices through
parallel contraction of two edges, see Fig.~\ref{fig:trees} 
\subref{fig:twocontract} for an example
($C_{1,2}: C=1,E=F=0$ in (\ref{eq.score})
test case $C_{ser}$ where we also allow for deletion of vertices in a
parent-child relationship (= serial)\footnote{Thereby we do not allow
  for deletion of grandparent-parent-child relationships, which
  preserves an efficient recurrence scheme.} without penalizing any
sort of deletion.  We achieved best results in the case of $C_{1,2}$
and further determined that to considerably penalize parallel
contraction in contrast to imposing only a relatively mild penalty for
isolated contraction yielded an optimal choice of parameters
$E=2,F=50$ (referred to as $C_{\rm Opt}$). We suggest ratios
$C/E=1/2,E/F=1/25$ as default settings. However, determination of
absolute values needs to put into context with orders of magnitude of
edge weights of the trees under consideration.\par As outlined in the
Methods section, inducing tree constraints considerably reduces the
search space, thereby allowing for an efficient and deterministic
method. To also highlight these effects, we further determine the size
of the largest correct mapping which does not violate the tree
constraints, $CP$ (``Constraint Positives'').  We compute
$RP=\frac{CP}{P}$ (``Relative Positives'') as the fraction of pairings
that can still be inferred, which is a value which reflects how the
reduction of search space influences the number of correct
pairings. We further compute ${\rm RelRec}=\frac{TP}{CP}$ (``Relative
Recall'') as a recall value which reflects how many of the correct
pairings possible were inferred by the algorithm in question. Note
that the heuristic search does not impose any constraints on the
search space hence $CP=P$ such that Recall and Relative Recall
coincide.  Juxtaposing $RP$ and RelRec values are meant to put usage
of tree topology into a general perspective. Moreover, RelRec values
certainly shed light on the effectiveness of the search strategy in
use.\par Table \ref{tab.overall} presents numbers of all $488$ tree
pairs. Following \cite{Izarzugaza08}, we also separate tree pairs
according to the numbers of leaves of the larger tree (see Table
\ref{tab.maxsize}) and the product of the numbers of leaves of the two
paired trees (see Table \ref{tab.space}) which, according to
\cite{Izarzugaza08}, quantifies search space size. Optimal values
are {\bf highlighted} in all categories.

\begin{table}[h] 
  \begin{center} {
      \begin{tabularx}{\columnwidth}{ C||C|C|C|C|C}
        Method & RP & Recall & RelRec & Precis & ${\rm F}_{0.25}$\\
        \hline\hline
        ${\rm C}_0$ & 0.546 & 0.330 & 0.557 & 0.447 & 0.438\\
        ${\rm C}_1$ & 0.610 & 0.378 & 0.586 & 0.475 & 0.468\\
        ${\rm C}_{\{1,2\}}$ & 0.612 & 0.377 & 0.581 & 0.471 & 0.464\\
        ${\rm C}_{ser}$ & 0.638 & 0.373 & 0.556 & 0.444 & 0.439\\
        ${\rm C}_{\rm Opt}$ & 0.612 & 0.380 & {\bf 0.588} & {\bf 0.479} & {\bf 0.472}\\
        \hline
        Heur. & {\bf 1.000} & {\bf 0.550} & 0.550 & 0.450 & 0.450
      \end{tabularx} }
    \caption{\label{tab.overall} Evaluation of our method with
      different choices of parameters and the previously published
      heuristic approach \cite{Izarzugaza08} (= Heur., values have
      been rounded to the order of $10^{-2}$). Baseline values for
      Recall and Precision are $\frac16=\approx 0.17$.}
  \end{center}
\end{table}

\begin{table}[h] 
  \begin{center} {
      \begin{tabularx}{\columnwidth}{C||CCC|CCC}
               & \multicolumn{6}{c}{MaxSize}\\
               & \multicolumn{3}{c|}{$<$120} & \multicolumn{3}{c}{$\ge$120} \\
        Meth. & Rec & Prec & ${\rm F}_{0.25}$ & Rec & Prec & ${\rm F}_{0.25}$ \\
               \hline
               \hline
        ${\rm C}_0$ & 0.389 & 0.511 & 0.502 & 0.200 & 0.305 & 0.296 \\
        ${\rm C}_1$ & 0.436 & 0.537 & 0.530 & 0.249 & 0.338 & 0.331 \\
        ${\rm C}_{1,2}$ & 0.436 & 0.534 & 0.527 & 0.245 & 0.331 & 0.527 \\
        ${\rm C}_{ser}$ & 0.437 & 0.525 & 0.519 & 0.231 & 0.262 & 0.260 \\
        ${\rm C}_{\rm Opt}$ & 0.439 & {\bf 0.541} & 0.534 & 0.251 & {\bf 0.340} & {\bf 0.333} \\
        \hline
        Heur. & {\bf 0.700} & {\bf 0.540} & {\bf 0.550} & {\bf 0.340} & 0.280 & 0.280 \\
      \end{tabularx} }
    \caption{\label{tab.maxsize} The comparison of our method with the
      heuristic search method shows favorable results for large trees
      ($\ge$120 leaves) for our method. For Heur., values have been
      rounded to the order of $10^{-2}$.}
  \end{center}
\end{table}

\begin{table}[h] 
  \begin{center} {
      \begin{tabularx}{\columnwidth}{C||CCC|CCC}
               &\multicolumn{6}{c}{Space} \\
               & \multicolumn{3}{c|}{$<$11680} & \multicolumn{3}{c}{$\ge$11680} \\
        Meth. & Rec & Prec & ${\rm F}_{0.25}$ & Rec & Prec & ${\rm F}_{0.25}$ \\
               \hline
               \hline
        ${\rm C}_0$ & 0.361 & 0.478 & 0.469 & 0.191 & 0.305 & 0.295\\
        ${\rm C}_1$ & 0.409 & 0.506 & 0.499 & 0.239 & 0.336 & 0.328\\
        ${\rm C}_{1,2}$ & 0.407 & 0.501 & 0.494 & 0.239 & 0.334 & 0.326 \\
        ${\rm C}_{ser}$ & 0.410 & 0.489 & 0.483 & 0.205 & 0.238 & 0.236 \\
        ${\rm C}_{\rm Opt}$ & 0.410 & {\bf 0.508} & 0.501 & 0.245 & {\bf 0.343} & {\bf 0.335}\\
        \hline
        Heur. & {\bf 0.640} & 0.500 & {\bf 0.510} & {\bf 0.280} & 0.20 & 0.20\\
      \end{tabularx} }
    \caption{\label{tab.space} The comparison of our method with the
      heuristic search method reveals favorable results for large
      search spaces (Space $\ge$11680). For Heur., values have been
      rounded to the order of $10^{-2}$.}
  \end{center}
\end{table}

\section{Discussion}

\paragraph{Runtime.}
The possibly most striking advantage of the topology-based approach is
the drastic reduction of runtime---we can align all trees in $\approx
1$ minute on a single processor laptop instead of $730$ hours on a
super computer. Note that there are rapidly growing large-scale
phylogenetic databases such as ENSEMBL \cite{Ensembl} or PhylomeDB
\cite{Phylome}, whose growth is further accelerated by
next-generation sequencing projects (as of $12$th August, 2011,
PhylomeDB contains $482,274$ phylogenetic trees). The reduction in
runtime delivered by our approach certainly overcomes a major
obstacle---we render large-scale mapping and, as a consequence,
comparison of paralog-rich gene trees feasible. Note that this
reduction has become possible by imposing both computationally and
biologically reasonable constraints on the search space while at the
same time allowing for an efficient scheme to find the global optimum
within these constraints.

\paragraph{Search Space Size / Recall.} 
Comparing $C_{\rm Opt}$ with the method of \cite{Izarzugaza08}
(Heuristic) overall, clearly, \cite{Izarzugaza08} achieve best
recall. As pointed out above, this comes as no surprise since we
cannot explore pairings that contradict the topologies of the paired
trees.  Quite surprisingly though, although usage of tree topology and
neighbor-joining trees in particular have been discussed rather
controversially \cite{Waddell07}, we find that still the majority of
pairings ($54.6\%$ with the strictest constraints and $61.2\%$ for
allowing isolated and parallel deletion) can be determined by a
topology-based approach. These numbers may put usage of
neighbor-joining tree topology in mirrortree approaches into a general
perspective.  Moreover, note that the fraction of correct domain pairs
computed by our method over that of the heuristic search method is
about $0.7$ ($=\frac{TP(C_{opt})}{TP(Heuristic)} = \frac{{\rm
    Recall}(C_{opt})}{{\rm Recall}(Heuristic)} = \frac{0.38}{0.55}$)
which is more than what was to be expected by reduction of the search
space ($\frac{CP(C_{opt})}{CP({\rm
    Heuristic})}=\frac{CP(C_{opt})}{P({\rm
    Heuristic})}=RP(C_{opt})=0.61$) which points out that we
compensate search space reduction by a more effective search
strategy. This becomes reflected by the better RelRec values of
$C_{opt}$.

\paragraph{Precision and F-Measure.}
Precision also favors the topology-based approach, at least on larger
(combinations of) trees (see column Prec in all three tables). Better
precision reflects a larger fraction of the correct domain pairs among
the pairs inferred overall and \cite{Tillier09} argue in a most recent
contribution that precision is more relevant than recall in mirrortree
approaches. Consequently, they suggest the F-measure
$F_{0.1}=\frac{(1+0.1^2)\cdot \Rec \cdot \Prec}{(0.1^2\Prec + \Rec)}$
to assess overall performance. We slightly take issue with this
suggestion as we feel that $F_{0.1}$ overrates Precision and instead
suggest the more balanced $F_{0.25}=\frac{(1+0.25^2)\cdot \Rec \cdot
  \Prec}{(0.25^2\Prec + \Rec)}$. We achieve better values in terms of
$F_{0.25}$ than \cite{Izarzugaza08} on pairs of larger trees. See Prec
and $F_{0.25}$ in tables \ref{tab.overall}, \ref{tab.maxsize} (in
particular $\ge 120$) and \ref{tab.space} (in particular $\ge 11680$)
for related results.

\paragraph{Conclusion.}

In summary, we have, for the first time, devised a deterministic and
efficient, polynomial-runtime mirrortree approach which directly
compares the gene trees, and not the distance matrices behind or
giving rise to them.  We have juxtaposed our approach with the most
recent, state-of-the-art matrix-based heuristic search procedure
without introducing further experimental biases.  Our tree
topology-based algorithm lists efficiency---its runtime is better by
several orders of magnitude, reducing runtime from several hundreds of
hours to only one minute when mirroring $\approx$500 trees---and
precision as its benefits. Recall is better for the heuristic search
which is explained by that the inherent search strategy does not
impose any constraints on the search space. Our advantages become most
obvious for large trees and in particular when both of the mirrored
trees are not small.  Here, our algorithm also achieves comparable
recall values while our advantages in precision become distinct. This
leads us to conclude that the heuristic method remains the better
choice for smaller trees and when runtime is not an issue. In case of
larger trees and in particular for large-scale studies, our approach
has considerable benefits. Note finally that we have been comparing
neighbor-joining which have been repeatedly exposed as suboptimal
choices of phylogenetic trees.  We believe that our approach can gain
from improvements in tree quality significantly more than the
matrix-based approaches.

\section*{Acknowledgment}

\paragraph{Funding.} 
Iman Hajirasouliha is supported by an NSERC Alexander Graham Bell
Canada Graduate Scholarships (CGS-D). Alexander Sch\"onhuth performed
most of this work while being supported by a postdoctoral fellowship
of the Pacific Institute of Mathematical Sciences. S.~Cenk Sahinalp
receives funds from the Natural Sciences and Engineering Research
Council of Canada (NSERC) and Bioinformatics for Combating Infectious
Diseases (BCID) for this project.

\bibliographystyle{plain}
\bibliography{genetree}

\end{document}